\documentclass[journal]{IEEEtran}
\usepackage{bm}
\usepackage{amssymb}
\usepackage{amsfonts}
\usepackage{balance}
\usepackage{booktabs}
\usepackage{amsmath}
\usepackage{tabularx}
\usepackage{subfigure}
\usepackage{graphicx}
\usepackage{multirow}
\usepackage{rotating}
\usepackage{algorithm}
\usepackage{url}
\usepackage{amsthm}
\usepackage{algcompatible}
\usepackage{epstopdf}
\usepackage{makecell}

\usepackage{color}
\begin{document}
\makeatletter
\newcommand{\rmnum}[1]{\romannumeral #1}
\newcommand{\Rmnum}[1]{\expandafter\@slowromancap\romannumeral #1@}
\makeatother
\title{Multi-party Secure Broad Learning System for Privacy Preserving}

\author{Xiao-Kai~Cao,~%\IEEEmembership{Student Member,~IEEE,}
 Chang-Dong~Wang,~\IEEEmembership{Member,~IEEE,}
 Jian-Huang~Lai,~\IEEEmembership{Senior Member,~IEEE,}\\
 Qiong~Huang,~\IEEEmembership{Member,~IEEE}
 C. L. Philip~Chen,~\IEEEmembership{Fellow,~IEEE}
 % <-this % stops a space
\thanks{The work was supported by National Key Research and Development Program of China (2021YFF1201200) and NSFC (61876193).}
\thanks{Xiao-Kai Cao, Chang-Dong Wang and Jian-Huang Lai are with School of Computer Science and Engineering, Sun Yat-sen University, Guangzhou, China, and also with Key Laboratory of Machine Intelligence and Advanced Computing, Ministry of Education, China.
  E-mail: caoxk@mail2.sysu.edu.cn,  changdongwang@hotmail.com, stsljh@mail.sysu.edu.cn.}
\thanks{Qiong Huang is with College of Mathematics and Informatics, South China Agricultural University, Guangzhou, China.
  E-mail: qhuang@scau.edu.cn.}
\thanks{C. L. Philip Chen is with School of Computer Science and Engineering, South China University of Technology, Guangzhou, China.
  E-mail: philip.chen@ieee.org.}
  \thanks{Corresponding author: Chang-Dong Wang.}
  }

% The paper headers
\markboth{IEEE Transactions on Cybernetics}%
{\MakeLowercase{\textit{Cao et al.}}: MSBLS}

\maketitle

\begin{abstract}
  Multi-party learning is an indispensable  technique for improving the learning performance via integrating data from multiple parties. Unfortunately, directly integrating multi-party data would not meet the privacy preserving requirements. Therefore, Privacy-Preserving Machine Learning (PPML) becomes a key research task in multi-party learning. In this paper, we present a new PPML method based on secure multi-party interactive protocol, namely Multi-party Secure Broad Learning System (MSBLS), and derive security analysis of the method. The existing PPML methods generally cannot simultaneously meet multiple requirements such as security, accuracy, efficiency and application scope, but MSBLS achieves satisfactory results in these aspects. It uses interactive protocol and random mapping to generate the mapped features of data, and then uses efficient broad learning to train neural network classifier. This is the first privacy computing method that combines secure multi-party computing and neural network. Theoretically, this method can ensure that the accuracy of the model will not be reduced due to encryption, and the calculation speed is very fast. We verify this conclusion on three classical datasets.
\end{abstract}


%\IEEEpeerreviewmaketitle
% this part is introduction
\section{Introduction}\label{sec: Introduction}
\IEEEPARstart{D}{ata} classification is a classical data analysis task, which is widely used in various fields. With the development of machine learning, various supervised neural network classifiers are proposed, along which the classification performance is significantly improved. On this basis, some attempts have been made to consider the task requirements in specific scenarios, in which privacy preserving is an invaluable requirement in recent years.  Take the medical scene as an example. Medical image classification is a typical task in medical data analysis \cite{DBLP:journals/tmi/GreenspanGS16}. However, the data samples of a single medical institution may not meet the needs of data analysis tasks \cite{DBLP:journals/hisas/RaghupathiR14}. For example, the data of tumor hospitals are mostly related to tumors. Due to the lack of data of other diseases, the trained classifier may diagnose non tumor patients as tumor patients; On the contrary, there are few various tumor data samples in general hospitals. It is a challenging issue to train a classifier from the relatively small number of high dimensional image data. In addition to the need for cooperation between specialized hospitals and general hospitals, small hospitals also need data help from large hospitals. A simple but effective strategy for addressing the issue is to integrate multi-party data from different medical image owners.

Unfortunately, directly integrating multi-party data would not meet the privacy preserving requirements. Medical data often contain patient identity information and health data, which belong to personal privacy information. If each medical institution directly fuses its patients' information for data analysis, the process will inevitably leak the private data to other medical institutions or the third-party server responsible for computing, which is extremely unsafe. Therefore, Privacy-Preserving Machine Learning (PPML) \cite{DBLP:conf/nips/ChaudhuriM08} becomes a key research task in medical image analysis. Compared with the traditional machine learning methods, PPML needs to consider security and communication cost. This means that it needs to find a balance between security, communication cost and model performance. In general, a complete PPML method is designed as follows. First, it encrypts the data in the local client, and then transmits the encrypted data to the central server. Finally, the central server conducts model training according to the algorithm design. In this process, more complex encryption methods may improve the security of data and reduce the performance of the model. In addition, some methods need to repeat the above steps many times in order to improve the model performance, which will increase the communication cost.

Broad Learning System (BLS) is a neural network model without deep structure~\cite{DBLP:journals/tnn/ChenL18, gong2021research}. BLS first generates a series of the mapped features by a large number of random transformations on the original data, then activates the random linear combination of the mapped features to obtain a series of enhancement features, and finally combines the mapped features and the enhancement features as the coefficient matrix of linear equations, the output class label as a non-homogeneous term, and calculates the weight coefficient (the solution of linear equations) through the approximate calculation method of pseudo inverse. The difference between BLS and deep neural network is that the deep neural network adjusts the weight coefficient through multiple iterations, so that the output result approximates the objective function. This process usually requires a long calculation time. BLS only needs to calculate the mapped features and the enhancement features once, and obtains the weight coefficients through a fast pseudo inverse approximation method. Although BLS will get some features with low contribution, it does not need to iterate repeatedly, so the consumption of computing resources is not high. In recent years, this method has been widely developed~\cite{DBLP:journals/tcyb/GanZCC22, DBLP:journals/tcyb/FengC20, DBLP:journals/tcyb/XuHCQ20} and used for solving various problems~\cite{DBLP:journals/tcyb/YeLC21, DBLP:journals/tcyb/HuangYC21, DBLP:journals/tcyb/GuoSLC21}.

PPML is a cross field involving machine learning and information security. Among them, cryptography is the theoretical cornerstone of information security. This means that ideal PPML needs to give full play to the advantages of machine learning in application and get the security guarantee of cryptography in theory. However, the focuses of these two areas are different. The former focuses more on the field of machine learning and application efficiency, while the latter focuses more on theory and security. Therefore, building a bridge between these two fields is a very important work. Although many efforts have been made on studying this problem, on the whole, most of the research work only focuses on application and efficiency, such as federal learning. The starting point of this paper is to find a way to closely combine these two fields. From the perspective of information security, this paper constructs an appropriate secure multi-party protocol to provide security protection for the clients involved in computing and achieve the security goal of the algorithm. From the perspective of machine learning, this paper uses the efficient broad learning system (broad neural network) to train the machine learning model. Theoretically, the broad learning system can approach any bounded function on any compact set, and its function is similar to that of deep neural network. Based on the above, we use random feature mapping to combine secure multi-party protocol and BLS. Among them, randomness protects the security of data from the perspective of information security, and the feature mapping can be used as the input information of BLS. Experiments show that this method will not lose the performance of the model on the premise of protecting data security. In addition, we analyze the security of this method, that is, the parties involved in the calculation cannot obtain the privacy data of other parties.

Aiming at the difficulties of the PPML method, this paper designs Multi-party Secure Broad Learning System (MSBLS) to implement privacy computing based on the technical characteristics of BLS. Specifically, the main ideas of this paper are as follows.

First of all, this paper encrypts data with the help of the generation process of the mapped features of BLS. BLS extracts random linear combination of the original features of data to generate the mapped features. We take random linear combination as the encryption process of data, so as to ensure the security of data. However, this is not a simple and direct task. We use the random coefficient matrix as the secret key, but different clients need to encrypt their own data with the same secret key to ensure that the mapped features are not chaotic (see Section~\ref{subsec: Privacy-preserving via random mapping} for detailed reasons). However, the shared secret key is likely to cause the disclosure of private data. Therefore, this paper designs an interactive protocol for encrypting data. This protocol can use the same secret key to encrypt the data of two clients and generate the mapped features. Afterwards, we simplify the generation process of the mapped features into a matrix form. The advantage of this method is that it can simplify the process of interactive protocol and make the protocol easier to be extended to different application scenarios. In the next part, we analyze the security of the computing process of the interactive protocol. The analysis results show that the protocol can guarantee the data security of two clients under the semi-honest model. In the end, MSBLS cleverly embeds the mapped features generated above into BLS and calculates model parameters. In order to verify the experimental performance of MSBLS, we conduct experiments on three classical data sets and simulate different actual scenarios. This includes the case where the number of data samples is unbalanced and the case where the distribution of data labels is non-uniform (see section~\ref{subsec: Datesets Description and Experimental Scenario} for a detailed description). Experiments show that even in extremely difficult scene tasks, compared with BLS, MSBLS can still protect the data security without losing the performance of the model. As a comparison, the current popular FedProx \cite{DBLP:conf/mlsys/LiSZSTS20} algorithm still has a small loss of accuracy. This means that MSBLS can simultaneously meet the requirements of security, communication cost and model performance.

In summary, the main contributions of this paper are as follows.
\begin{enumerate}
\item According to the requirements of PPML, this paper proposes the MSBLS method, which inherits the advantages of Secure Multi-party Computing (SMC) and BLS. That is, we design an interactive protocol to encrypt the data of the two clients and generate the mapped features, and use the mapped features to calculate the machine learning model parameters.

\item This paper simplifies the generation of the mapped features in BLS, making the process of interactive protocol more concise.

\item This paper gradually analyzes whether the data received by each client and server can recover the original data. The security analysis results show that the amount of information they hold is not enough to recover the original data.
\end{enumerate}

The rest of this paper is as follows. Section~\ref{sec: Related Work} introduces several commonly used PPML methods and points out the advantages of MSBLS. Section~\ref{sec: BLS} introduces the neural network classifier namely BLS used in this paper. Section~\ref{sec: Proposed Method} describes the MSBLS method in detail, including the problems to be solved, the simplified way of BLS and the specific solutions to the problems. Section~\ref{Security analysis} analyzes the security of the proposed scheme. Section~\ref{sec: Experiments} designs the experimental process and analyzes the experimental results. Section~\ref{sec: Conclusions} summarizes the work of this paper and discusses the possible future research work.

\section{Related Work}\label{sec: Related Work}

To solve the problems of security and computing efficiency in the process of privacy computing, many efforts have been made in developing different solutions from different perspectives \cite{DBLP:journals/tcyb/LouYWY18, DBLP:journals/tcyb/YangY22}. Among them, differential privacy \cite{DBLP:conf/tamc/Dwork08}, homomorphic encryption \cite{rivest1978data}, federated learning \cite{DBLP:journals/corr/KonecnyMRR16, DBLP:journals/corr/KonecnyMYRSB16, DBLP:journals/corr/McMahanMRA16} and secure multi-party computing \cite{DBLP:conf/focs/Yao82b} are the fastest-growing methods.

The concept of differential privacy was first proposed by Microsoft in 2006~\cite{dwork2006differential}. The differential privacy method protects the real value of sensitive data by adding noise to the data, and uses the corresponding method to obtain the real value in the query process~\cite{DBLP:conf/tamc/Dwork08}. The advantage and disadvantages of this method coexist. Although adding noise protects the security of data, it also destroys the original characteristics of data to a certain extent. Therefore, there is no ideal differential privacy algorithm that can equilibrate the relationship between security, computational efficiency and query accuracy.

Homomorphic encryption was proposed by Rivest et al. in 1978 \cite{rivest1978data}. In 2009, Gentry improved homomorphic encryption to obtain fully homomorphic encryption \cite{gentry2009fully}. The method of homomorphic encryption is to find a mapping function to encrypt the data, and then use the encrypted data for analysis, which ensures that the analysis result is the same as that obtained by directly using the original data for analysis. However, the efficiency of fully homomorphic encryption is low, and the encrypted data can only use addition and multiplication (the use of square root operation or other operations will lead to inconsistent results), so homomorphic encryption can only be applied to simple data analysis methods such as linear model \cite{DBLP:journals/corr/AslettEH15, DBLP:conf/kdd/LiH20, DBLP:conf/acns/GiacomelliJJPY18}.

Federal learning is a privacy-preserving machine learning framework proposed by Google in 2016 \cite{DBLP:journals/corr/KonecnyMRR16, DBLP:journals/corr/KonecnyMYRSB16, DBLP:journals/corr/McMahanMRA16}. In recent years, it has been widely studied and applied to various tasks \cite{DBLP:journals/spm/LiSTS20, DBLP:journals/corr/abs-2006-02931, DBLP:journals/corr/abs-1908-07873, DBLP:journals/corr/abs-2108-13323, DBLP:journals/corr/abs-2102-04925, DBLP:journals/corr/abs-2109-05446, DBLP:journals/tcyb/LeLMZZL21}. In 2017, McMahan et al. proposed FedAvg \cite{DBLP:conf/aistats/McMahanMRHA17}. Its basic steps are as follows: first, the central server sends the initial model parameters (global model parameters) to each client, then each client trains the local model parameters with local data, and then sends the updated model parameters to the central server for aggregation to obtain new global model parameters. The above steps are repeated iteratively until the global model parameters converge. In 2020, Li et al. generalized and reparameterized FedAvg to obtain FedProx \cite{DBLP:conf/mlsys/LiSZSTS20}. Despite the success, there is a lack of the relevant literatures for proving the security of federal learning. Moreover, although federated learning only transmits model parameters rather than the original data, there are literatures that use the model parameters of federated learning to restore the original data \cite{DBLP:conf/nips/GeipingBD020, DBLP:journals/corr/abs-2110-15122}.

Secure Multi-party Computing (SMC) is a privacy protection computing method proposed by Yao for the millionaire problem in 1982 \cite{DBLP:conf/focs/Yao82b}. For a group of participants participating in the calculation, each participant has its own private data and does not trust other participants or any third party. On this premise, SMC constructs an interactive protocol to calculate a target result from multi-party data. The main difficulty of the SMC protocol is to construct corresponding protocols for different tasks. In 2014, Bost et al. \cite{DBLP:conf/ndss/BostPTG15} constructed three SMC protocols for classification problems, which are respectively used to realize the privacy constraints of hyperplane decision problem, medium vector Bayesian problem and decision tree problem. Cock et al. proposed another decision tree privacy protection protocol in 2018 \cite{DBLP:journals/tdsc/CockDHKNPT19}. In 2020, Pan et al. proposed a joint feature selection algorithm based on the SMC interactive protocol \cite{pansecure}. Despite the long-term development, the SMC methods can only be applied to the traditional linear machine learning models. At present, there is still a lack of neural network based models with privacy preserving for secure multi-party computing.

In this paper, we propose a new privacy-preserving machine learning framework called Multi-party Secure Broad Learning System (MSBLS). Specifically, we construct a neural network classifier based on SMC, which inherits the powerful fitting ability of BLS with the secure computing ability of SMC to form an efficient and secure privacy computing framework. It is especially suitable for integrating multi-party image data with preserving privacy.

Compared with the above methods, the proposed MSBLS method has the following advantages.

\begin{itemize}
    \item Compared with differential privacy, MSBLS can make full use of the original data information and has higher security. It can meet the needs of security, computing efficiency and query accuracy.

    \item Compared with homomorphic encryption, MSBLS can achieve the same effect of homomorphic encryption, but it will not be constrained by homomorphic addition and multiplication. In other words, MSBLS can be applied to a wider range of classification tasks, and the computational efficiency is very high.

    \item Compared with federated learning, MSBLS also uses neural network as classifier for model training, which can make full use of the advantages of machine learning methods. The difference is that MSBLS does not require a large number of iterations, so it can reduce the traffic and computing time. In addition, MSBLS provides security analysis, which theoretically ensures the data security of the client participating in the calculation.

    \item Compared with other SMC methods, MSBLS is not limited to the traditional linear machine learning methods, but uses the neural network method with stronger classification and generalization ability as the classifier.
\end{itemize}

\section{Broad Learning System}\label{sec: BLS}

BLS is a neural network model for supervised machine learning without deep structure. Compared with the traditional deep neural network, BLS uses the transverse broad structure to establish the model framework, and does not require iterative updating procedure to achieve a very good training effect. In this section, we will introduce the basic structure of BLS.

\begin{figure}[!t]%
\centering
\includegraphics[width=0.4\textwidth]{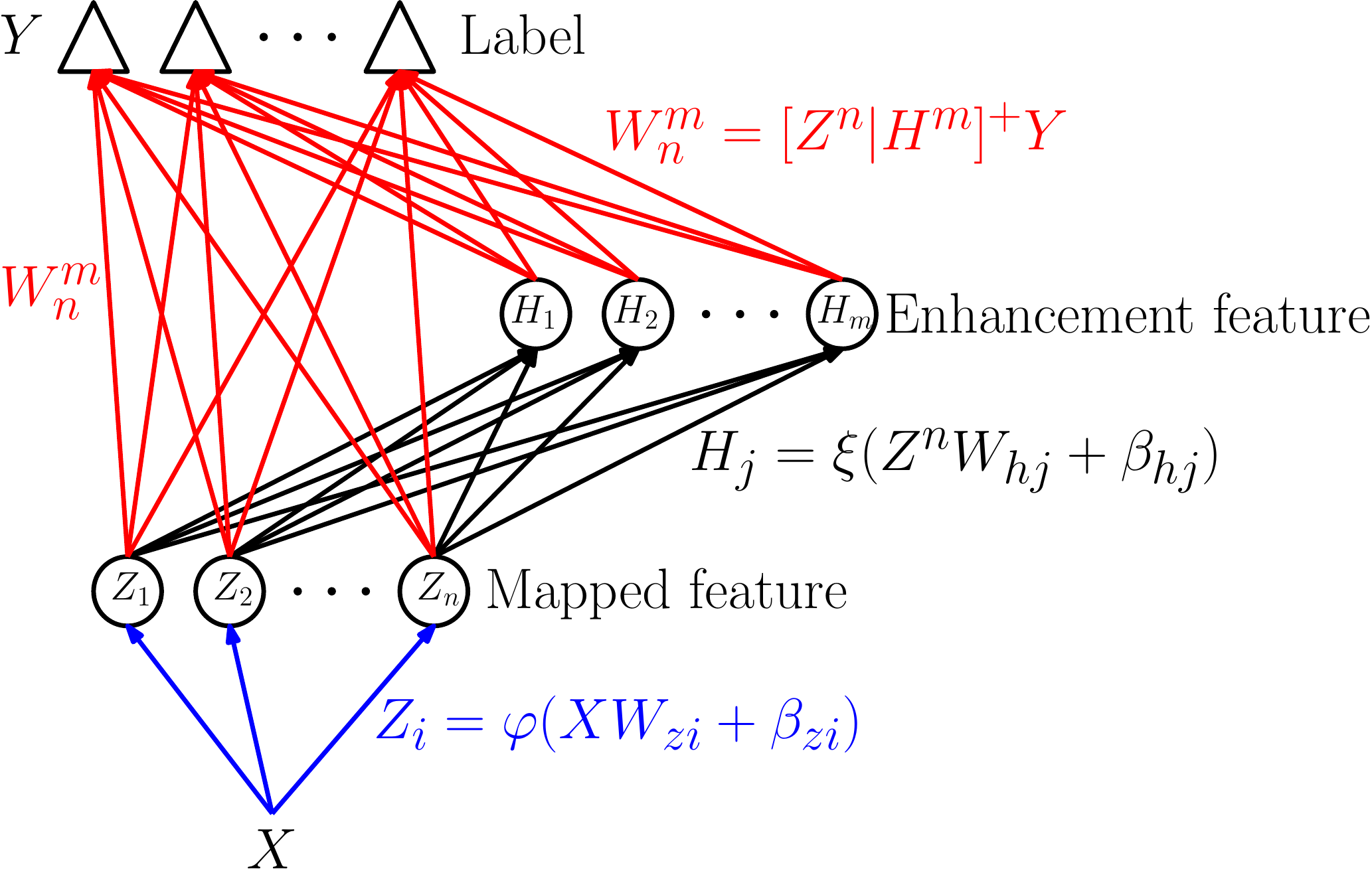}\\
\caption{The BLS framework.}\label{figure_BLS}
\end{figure}

As shown in \figurename~\ref{figure_BLS}, first, BLS uses a set of random feature maps to generate $n$ mapped features of data samples $\mathbf{X}\in\mathbb{R}^{N\times d}$, that is,
\begin{eqnarray}\label{Z_i}
\mathbf{Z}_i = \varphi (\mathbf{X} \mathbf{W}_{zi} + \bm{\beta}_{zi}), i=1,2,\ldots, n,
\end{eqnarray}
where $N$ and $d$ are the number and dimension of data samples, $d_z$ is the dimension of each mapped feature, $\mathbf{W}_{zi}\in\mathbb{R}^{d\times d_z}$ is the randomly generated weight matrix, and $\bm{\beta}_{zi}\in\mathbb{R}^{N\times d_z}$ is the randomly generated bias matrix with the same row, i.e., for each $i$, $\bm{\beta}_{zi}(j,:) = \bm{\beta}_{zi}(1,:), \forall j=1,2,\dots,N$.

Then, BLS uses a set of nonlinear activation functions to act on the mapped features $\mathbf{Z}^n$ to generate $m$ enhancement features, that is,
\begin{eqnarray}\label{H_j}
\mathbf{H}_j = \xi (\mathbf{Z}^n \mathbf{W}_{hj} + \bm{\beta}_{hj}), j=1,2,\ldots, m,
\end{eqnarray}
where $\mathbf{W}_{hi}\in\mathbb{R}^{nd_z\times d_h}$ is the randomly generated weight matrix, $\bm{\beta}_{hi}\in\mathbb{R}^{N\times d_h}$ is the randomly generated bias matrix with the same row, and $d_h$ is the dimension of each enhancement feature. The activation function $\xi(\cdot)$ can be set as the commonly used tangent function or sigmoid function.

Finally, BLS maps the concatenation of the mapped features and the enhancement features to the predicted output label vector $\mathbf{\hat{Y}}\in\mathbb{R}^{N\times d_Y}$ by means of a learnable weight matrix  $\mathbf{W}^m_n\in\mathbb{R}^{(nd_z+md_h)\times d_Y}$ as follows,
\begin{eqnarray}\label{Y}
\mathbf{\hat{Y}} &=& [\mathbf{Z}_1\mid \mathbf{Z}_2\mid \ldots \mid\mathbf{Z}_n\mid  \mathbf{H}_1\mid \mathbf{H}_2\mid \ldots \mid\mathbf{H}_m]\mathbf{W}^m_n\nonumber\\
&=& [\mathbf{Z}^n\mid \mathbf{H}^m]\mathbf{W}^m_n\nonumber\\
&=& \mathbf{A} \mathbf{W}^m_n,
\end{eqnarray}
where $d_Y$ is the dimension of the output label vectors. And in the training procedure, $\mathbf{W}^m_n$ can be solved by the approximate solution method of matrix pseudo inverse, i.e.,
\begin{equation}
 \mathbf{W}^m_n=\mathbf{A}^+\mathbf{Y}
\end{equation}
where
\begin{eqnarray}\label{A_pinv}
\mathbf{A}^+ = \lim\limits_{\lambda \rightarrow 0} (\lambda \mathbf{I}+\mathbf{A} \mathbf{A}^{\top})^{-1}\mathbf{A}^{\top},
\end{eqnarray}
where $\mathbf{Y}\in\mathbb{R}^{N\times d_Y}$ is the ground-truth class label matrix with each row being a one-hot vector representing the ground-truth class label of the corresponding sample.
For other calculation details of BLS, readers can read \cite{DBLP:journals/tnn/ChenL18, gong2021research}.

\section{Proposed Method}\label{sec: Proposed Method}

In this section, firstly, we describe the privacy computing problems that need to be solved. Secondly, we briefly summarize the MSBLS framework proposed in this paper. Thirdly, we simplify the generation of the mapped features in BLS. Fourthly, we analyze in detail the necessity for different clients to use the same secret key and propose an interactive protocol to solve this problem. Finally, we describe the proposed MSBLS method.

\subsection{Problem Statement}

In this section, we assume that two clients hold privacy data and randomly generate some secret keys. A third-party server assists the calculation. To protect data security, all unencrypted private data and secret keys cannot leave the client. The purpose of the method we designed is to calculate the mapped features in Eq.~\eqref{Z_i} required by BLS under the above premise, and finally calculate the parameters of the machine learning model of the neural network classifier.

\subsection{Overview of the MSBLS Framework}

\begin{figure}[!t]%
\centering
\includegraphics[width=0.4\textwidth]{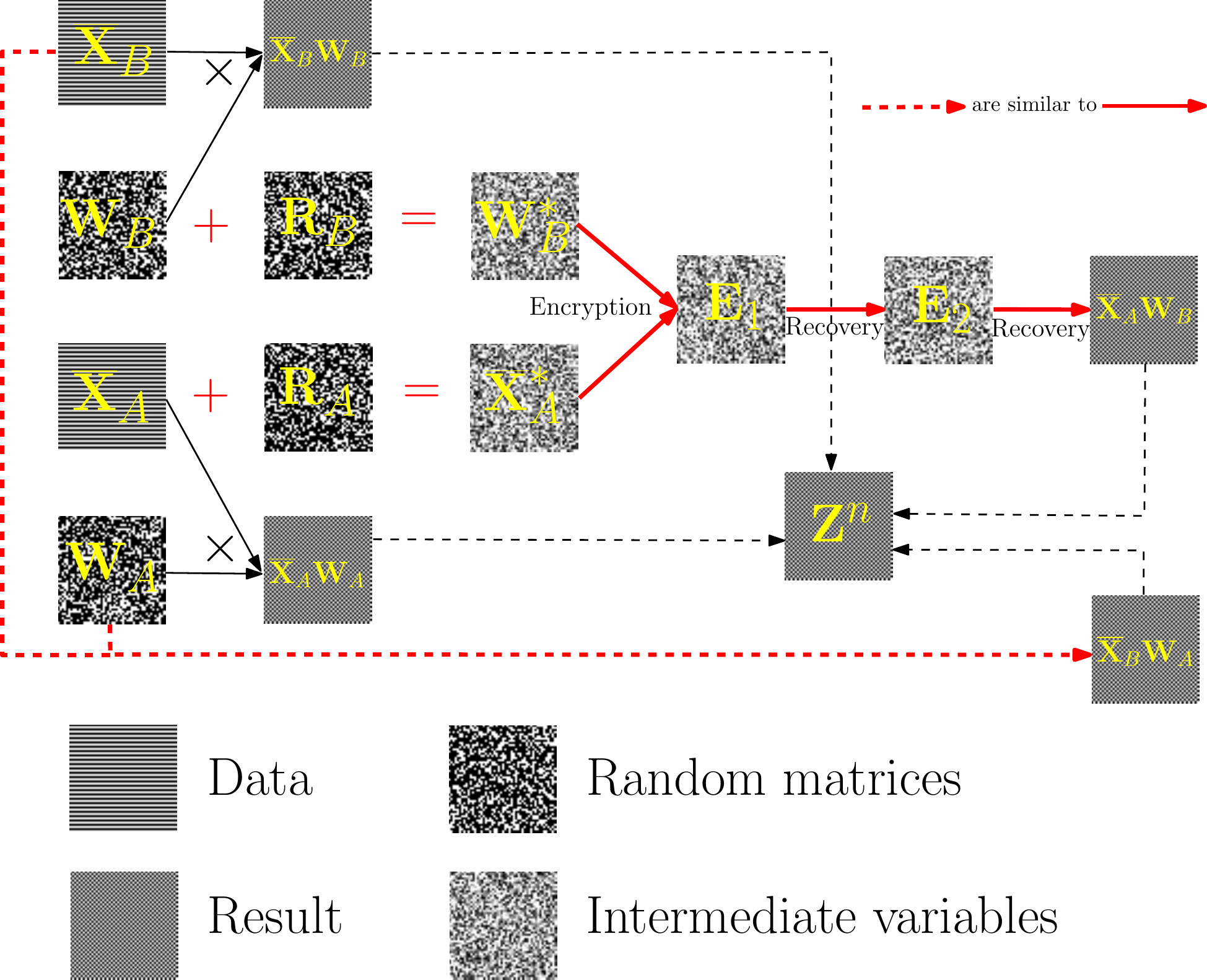}\\
\caption{Interactive protocol framework.}\label{Interative_Protocol_Framework}
\end{figure}

In this paper, we propose a privacy preserving multi-party machine learning method. It is very efficient and suitable for a variety of scenarios, which does not lose the information of data features while ensuring data security. Specifically, we use the secure multi-party computing protocol to protect the input data of the broad learning system, and calculate the mapped feature of the data in a secure computing environment. The computing idea of the protocol is shown in \figurename~\ref{Interative_Protocol_Framework}. In theory, the protocol can ensure that the trained machine learning model has the same performance as the machine learning model trained by direct fusion data. And, the experimental results confirm this conclusion. Moreover, different from homomorphic encryption, this method only needs few computing resources and the training speed is very fast.

\begin{figure}[!t]%
\centering
\includegraphics[width=0.4\textwidth]{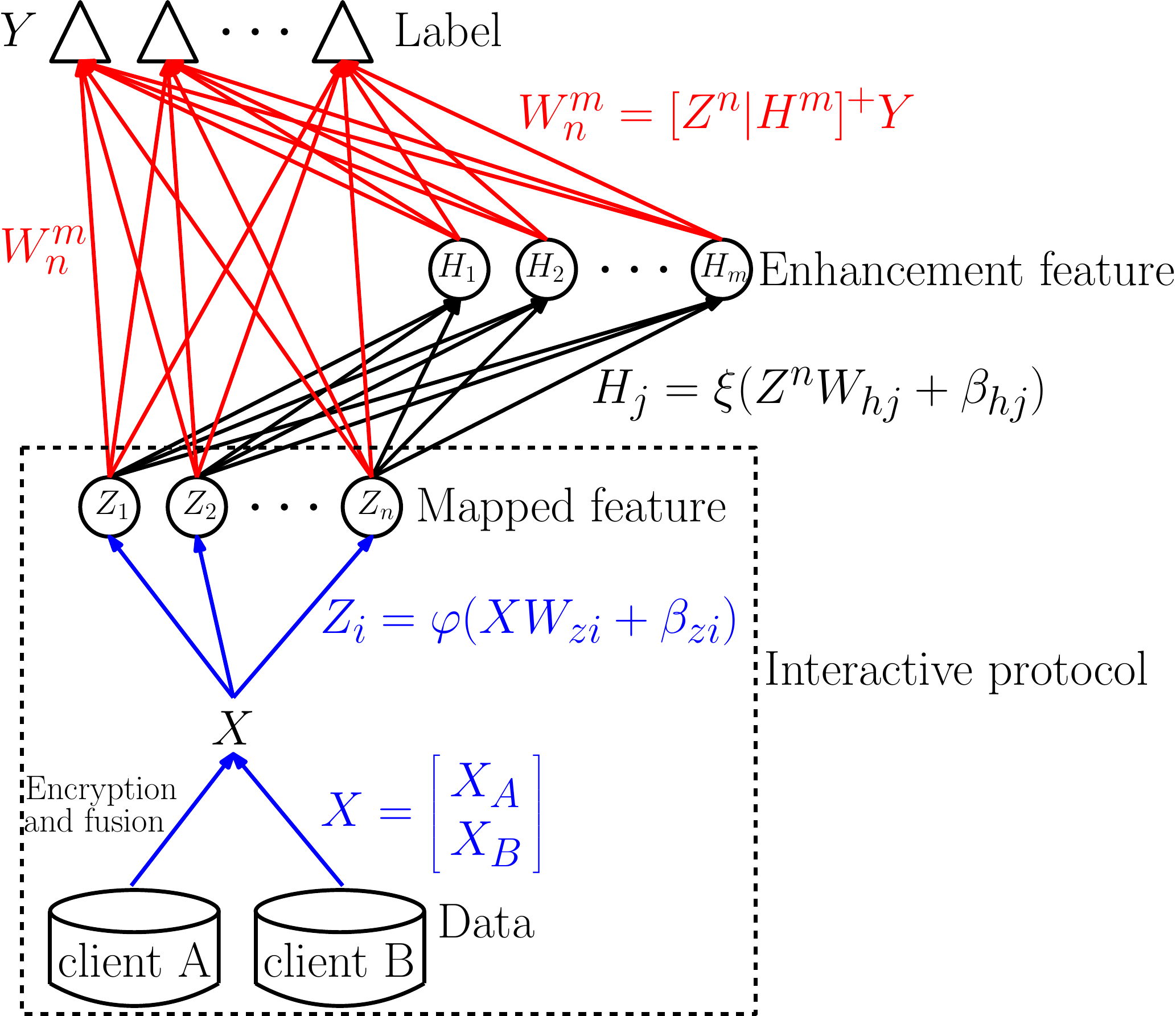}\\
\caption{The MSBLS framework.}\label{MSBLS_framework}
\end{figure}

First, MSBLS performs feature fusion on the input data of the two clients while preserving privacy. Specifically, MSBLS uses a third-party server to assist two clients in multiple rounds of interactive communication. The purpose of this process is to encrypt data and generate the mapped features. It is worth mentioning that neither two clients nor the third-party server can recover data that does not belong to them in this process. Then, in order to mine the nonlinear features of the data, similar to BLS, MSBLS activates the mapped feature to generate the enhancement feature. In this process, the original data will not be used, so the privacy of data is preserved. Finally, the mapped features and the enhancement features are concatenated to form a feature matrix, a linear mapping is established between the feature matrix and labels, and the model parameters are solved  by calculating the pseudo inverse of the matrix. Figure~\ref{MSBLS_framework} illustrates the MSBLS framework.

Theoretically, this encryption method will not lose the intrinsic information of the original data, so it can ensure that the trained model can generate the same results as applying the classical BLS on the direct fusion data, and our experiments on three classical image classification datasets confirm the above theoretical analysis. In addition, we analyze that in this process, the two clients providing data and the third-party server assisting the computation cannot recover the original data that does not belong to them. Moreover, the number of communications required by this method to process data of any scale is constant (12 times), which means that it is difficult to recover data through statistical law (law of large numbers). Finally, the method uses BLS as the neural network classifier, and its training speed is much faster than that of deep neural network, which ensures the high efficiency of the method.

\subsection{Simplification of BLS}

The MSBLS designed in this paper needs to send encryption features to achieve privacy computing. Since a large number of the mapped features need to be calculated in Eq.~\eqref{Z_i}, this process needs to be simplified to facilitate the design of interactive protocols.

In theory, the activation function $\varphi(\cdot)$ has no specific requirements in Eq.~\eqref{Z_i}. In this paper, in order to simplify the model and reduce the number of communications, we use linear function as $\varphi(\cdot)$. In addition, functions such as convolution function and nonlinear function can also be used.
Since $\varphi (\cdot)$ is a linear function, Eq.~\eqref{Z_i} can be re-written as follows,
\begin{small}
\begin{eqnarray}\label{Z_n}
&&\mathbf{Z}^n \nonumber\\
&=& [\mathbf{Z}_1 \mid \mathbf{Z}_2 \mid \ldots \mid \mathbf{Z}_n]\nonumber\\
&=&\big[\varphi (\mathbf{X} \mathbf{W}_{z1} + \bm{\beta}_{z1})\mid \varphi (\mathbf{X} \mathbf{W}_{z2} + \bm{\beta}_{z2})\mid \ldots \mid \nonumber\\
&&\varphi (\mathbf{X} \mathbf{W}_{zn} + \bm{\beta}_{zn})\big]\nonumber\\
&=&\varphi \big([\mathbf{X} \mathbf{W}_{z1} + \bm{\beta}_{z1}\mid \mathbf{X} \mathbf{W}_{z2} + \bm{\beta}_{z2}\mid \ldots \mid \mathbf{X} \mathbf{W}_{zn} + \bm{\beta}_{zn}]\big)\nonumber\\
&=&\varphi\Bigg(
\Bigg[
\left[\begin{array}{c c}
\mathbf{X} & \mathbf{1}^{N\times 1}
\end{array}\right]\times
\left[\begin{array}{c}
\mathbf{W}_{z1} \\
\bm{\beta}_{z1}(1,:)
\end{array}\right]\mid
\left[\begin{array}{c c}
\mathbf{X} & \mathbf{1}^{N\times 1}
\end{array}\right]\times\nonumber\\
&&
\left[\begin{array}{c}
\mathbf{W}_{z2} \\
\bm{\beta}_{z2}(1,:)
\end{array}\right]
\mid\cdots\mid
\left[\begin{array}{c c}
\mathbf{X} & \mathbf{1}^{N\times 1}
\end{array}\right]\times
\left[\begin{array}{c}
\mathbf{W}_{zn} \\
\bm{\beta}_{zn}(1,:)
\end{array}\right]
\Bigg]
\Bigg)\nonumber\\
&=&\varphi\Bigg(
\left[\begin{array}{c c}
\mathbf{X} & \mathbf{1}^{N\times 1}
\end{array}\right]\times \nonumber\\
&&
\left[\begin{array}{cccc}
\mathbf{W}_{z1} & \mathbf{W}_{z2} & \cdots & \mathbf{W}_{zn} \\
\bm{\beta}_{z1}(1,:) & \bm{\beta}_{z2}(1,:) & \cdots & \bm{\beta}_{zn}(1,:)
\end{array}\right]
\Bigg)\nonumber\\
&=&\varphi (\overline{\mathbf{X}} \times \mathbf{W}),
\end{eqnarray}
\end{small}
where $\mathbf{1}^{N\times 1}$ is a column vector of dimension $N$ with all elements being $1$, and $\overline{\mathbf{X}} = [\mathbf{X} ~~\mathbf{1}^{N\times 1}]\in\mathbb{R}^{N\times (d+1)}$. Since $\mathbf{W}_{zi}, i=1,2,\ldots, n$ are random matrices and $\bm{\beta}_{zi}(1,:),i=1,2,\ldots, n$ are random vectors, they can be written in the form of random matrix
\begin{equation}
	\mathbf{W} = \left[\begin{array}{cccc}
		\mathbf{W}_{z1} & \mathbf{W}_{z2} & \cdots & \mathbf{W}_{zn} \\
		\bm{\beta}_{z1}(1,:) & \bm{\beta}_{z2}(1,:) & \cdots & \bm{\beta}_{zn}(1,:)
	\end{array}\right]\in\mathbb{R}^{(d+1)\times nd_z}.
\end{equation}
Without loss of generality, the random matrix $\mathbf{W}$ can be expressed by the product of two random matrices $\mathbf{W}_0\in\mathbb{R}^{(d+1)\times nd_z}$ and $\mathbf{W}_1\in\mathbb{R}^{nd_z\times nd_z}$ (refer to Section \ref{Security analysis} for the necessity of this step), i.e.
\begin{eqnarray}\label{Z_n_3}
\mathbf{Z}^n = \varphi (\overline{\mathbf{X}} \mathbf{W}_0 \mathbf{W}_1).
\end{eqnarray}

In addition, the goal of Eqs.~\eqref{Z_n}-\eqref{Z_n_3} is to make the interactive protocol in Algorithm~\ref{Interactive protocol of MSBLS} (as will be described later) more concise and easy to read and greatly reduce the number of communications. Readers can change $\varphi(\cdot)$ into a nonlinear function according to their own needs. On this basis, the protocol in Algorithm~\ref{Interactive protocol of MSBLS} is still applicable.

\subsection{Privacy-preserving via Random Mapping}\label{subsec: Privacy-preserving via random mapping}

In this section, we will introduce the privacy-preserving part of MSBLS in detail.

\begin{figure}[!t]%
\centering
\includegraphics[width=0.4\textwidth]{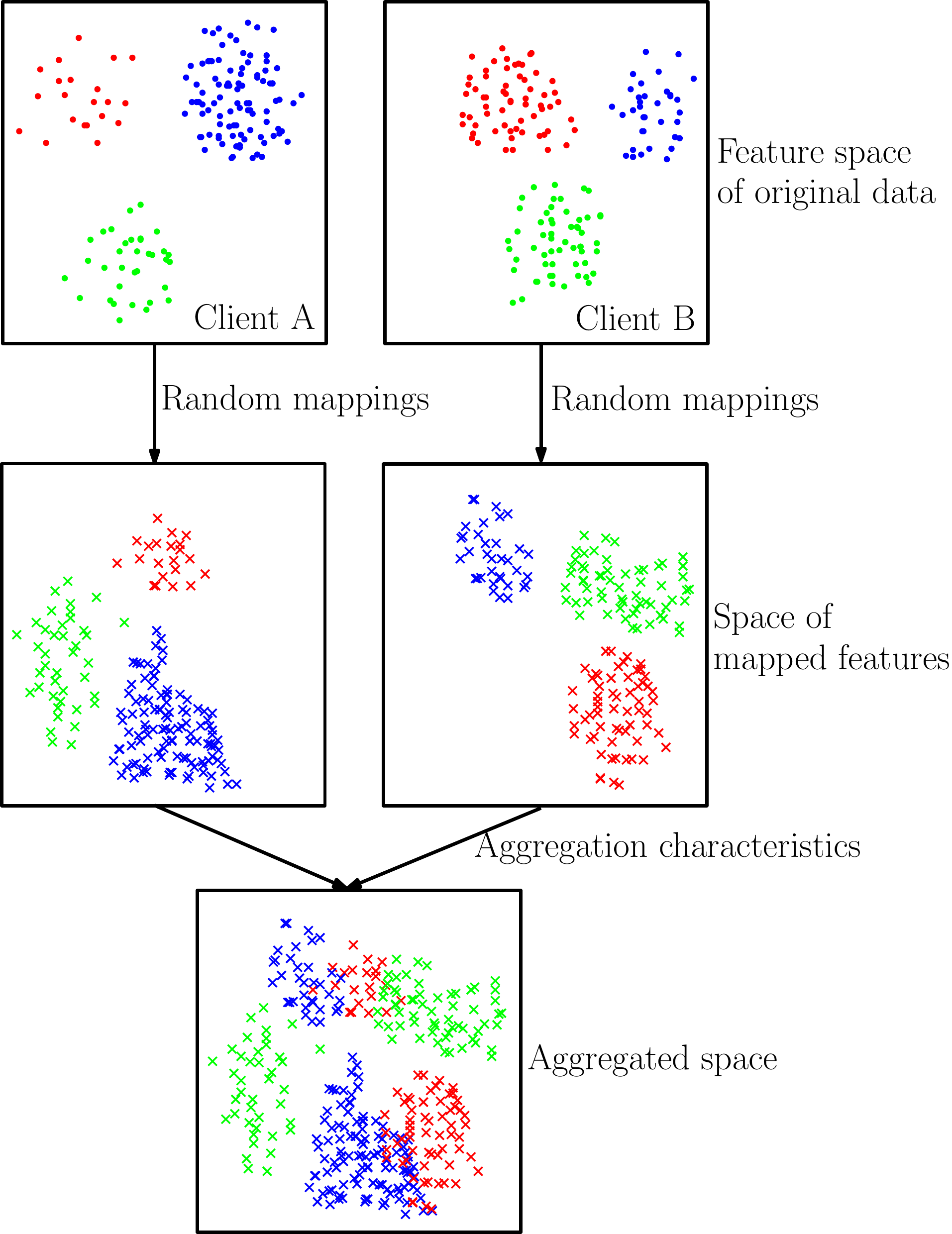}\\
\caption{Disadvantages of using different random mappings.}\label{figure_random_mapping}
\end{figure}

%We assume that two clients need to encrypt, fuse and analysis the data, and another third-party server assists in calculation. Let client A and client B hold data $\mathbf{X}_A$ and $\mathbf{X}_B$ respectively. Similar to equation \eqref{Z_n_3}, we use random mapping to generate data features, that is
Assume that we need to encrypt the data from two clients A and B, and then integrate and analyze ciphertext. In addition, there is a third-party server to assist the calculation. The implementation of data encryption generally needs a secret key. A common key encryption method is to encrypt data by using random mapping (secret key). However, if the data encrypted by random mapping is directly used in training machine learning models, it may cause performance loss, i.e., the original features are largely damaged by the random mapping. Therefore, we need to make random mapping as part of model training, which seems contradictory, but in fact, this is not an impossible work. For example, the deep neural network will use the random initial weight coefficient, the evolutionary algorithm will use the random initial population, and some clustering algorithms will carry out the random initial classification of samples. However, the  aforementioned randomness is only regarded as initialization which will be updated in the later steps and hence cannot be considered as data encryption. In this paper, we use the random mapping of BLS to achieve this purpose. Specifically, according to the previous description, BLS uses random mapping to generate the mapped features of data. From the perspective of data security, this process can be regarded as encrypting data. Therefore, using BLS can integrate data encryption and model training, and avoid the loss of model performance caused by encryption. However, it is not a trivial task. Specifically, if clients A and B directly use a set of confidential random mapping to encrypt data and aggregate the results as the mapped features in Eq.~\eqref{Z_i}, the feature extraction methods of the two clients will be inconsistent.
\begin{itemize}
    \item If A and B hold the same samples, using different random mappings will generate different mapped features, which is obviously not conducive to the subsequent model training.
    \item If A and B hold different samples, using different random mappings may generate the same mapped features, such as $(1,2,3)\times(2,1,0)^{\top}=(2,2,1)\times(0.5,0.5,2)^{\top}$, which is obviously not conducive to the subsequent model training.

    \item Generally, we hope that in the data sample space, the data distribution of the same label will be concentrated as much as possible, and the data distribution of different labels will be dispersed as much as possible. However, if two clients use different random mappings to generate the mapped features and jointly participate in the next operation, the mapped features of different labels may significantly overlap, and the mapped features of the same label may be distributed in multiple centralized areas. As shown in \figurename~\ref{figure_random_mapping}, this will seriously affect the training of model parameters.
\end{itemize}
On the contrary, if clients A and B use the same random mapping to encrypt the data respectively, it is equivalent to that they have obtained each other's secret key (random mapping), which is extremely unsafe for the security of the data. To sum up, assume that client A and client B hold data $\mathbf{X}_A\in\mathbb{R}^{N_A\times d}$ and $\mathbf{X}_B\in\mathbb{R}^{N_B\times d}$ respectively, with $N_A$ and $N_B$ being the number of samples of clients A and B, and $N=N_A+N_B$. We use the following methods for data encryption and feature extraction,
\begin{eqnarray}\label{Z_n_4}
\mathbf{Z}^n &=& \varphi (\overline{\mathbf{X}} \mathbf{W}_0 \mathbf{W}_1)\nonumber\\
&=& \varphi \bigg(
\bigg[\begin{array}{l}
\overline{\mathbf{X}}_A\\
\overline{\mathbf{X}}_B
\end{array}\bigg] \times
\big[\begin{array}{l}
\mathbf{W}_A ~~ \mathbf{W}_B
\end{array}\big] \times \mathbf{W}_1\bigg)\nonumber\\
&=& \varphi \bigg(
\bigg[\begin{array}{l}
\overline{\mathbf{X}}_A \mathbf{W}_A ~~~\overline{\mathbf{X}}_A \mathbf{W}_B\\
\overline{\mathbf{X}}_B \mathbf{W}_A ~~~\overline{\mathbf{X}}_B \mathbf{W}_B
\end{array}\bigg] \times \mathbf{W}_1\bigg),
\end{eqnarray}
%\begin{eqnarray}\label{Z_n_5}
%\mathbf{Z}^n = \varphi \bigg(
%\bigg[\begin{array}{l}
%\overline{\mathbf{X}}_A \times \mathbf{W}_A\\
%\overline{\mathbf{X}}_B \times \mathbf{W}_B
%\end{array}\bigg]\bigg),
%\end{eqnarray}
where $\overline{\mathbf{X}}_A= [\mathbf{X}_A ~~\mathbf{1}^{N_A\times 1}]\in\mathbb{R}^{N_A\times (d+1)}$ and $\overline{\mathbf{X}}_B= [\mathbf{X}_B ~~\mathbf{1}^{N_B\times 1}]\in\mathbb{R}^{N_B\times (d+1)}$ are the augmented matrices of $\mathbf{X}_A$ and $\mathbf{X}_B$ respectively, $\mathbf{W}_A\in\mathbb{R}^{(d+1)\times \frac{nd_z}{2}}$ and $\mathbf{W}_B\in\mathbb{R}^{(d+1)\times \frac{nd_z}{2}}$ are two random matrices, and $\mathbf{W}_1\in\mathbb{R}^{nd_z\times nd_z}$. In order to calculate the mapped features using Eq.~\eqref{Z_n_4}, it is necessary to calculate four block matrices $\overline{\mathbf{X}}_A \mathbf{W}_A$, $\overline{\mathbf{X}}_A \mathbf{W}_B$, $\overline{\mathbf{X}}_B \mathbf{W}_A$ and $\overline{\mathbf{X}}_B \mathbf{W}_B$.

Firstly, the random matrices $\mathbf{W}_A$ and $\mathbf{W}_B$ are generated by clients A and B respectively, and therefore $\overline{\mathbf{X}}_A \mathbf{W}_A$ and $\overline{\mathbf{X}}_B \mathbf{W}_B$ can be directly calculated by the two clients respectively. Furthermore, in order to calculate $\overline{\mathbf{X}}_A \mathbf{W}_B$ and $\overline{\mathbf{X}}_B \mathbf{W}_A$ in a secure environment, a third-party server needs to be introduced to assist the calculation. The interactive protocol framework for calculating $\overline{\mathbf{X}}_A \mathbf{W}_B$ is illustrated in \figurename~\ref{Interative_Protocol_Framework}. After generating the four block matrices, the mapped features are generated by the third-party server.

For clarity, Algorithm~\ref{Interactive protocol of MSBLS} summarizes the detailed generation process of complete data encryption and the mapped features, where $\mathbf{R}_A\in\mathbb{R}^{N_A\times (d+1)}$, $\mathbf{R}_B\in\mathbb{R}^{(d+1)\times \frac{nd_z}{2}}$ and $\mathbf{R}_b\in\mathbb{R}^{N_A\times \frac{nd_z}{2}}$ are three intermediate variables for encryption, and $\mathbf{E}_1\in\mathbb{R}^{N_A\times \frac{nd_z}{2}}$ and $\mathbf{E}_2\in\mathbb{R}^{N_A\times \frac{nd_z}{2}}$ are two encrypted intermediate variables.

% After generating the mapped features by using Algorithm~\ref{Interactive protocol of MSBLS}, formulas \eqref{H_j} - \eqref{A_pinv} are used to calculate the model parameters of BLS.

\begin{algorithm}[!t]%%\vskip -0.06in
\caption{The data encryption and mapped feature generation procedures of MSBLS.}
\label{Interactive protocol of MSBLS}
{\bfseries Input:} Data: Client A: $\mathbf{X}_A$, Client B: $\mathbf{X}_B$, feature mappings: $\varphi$, number of mapped features: $n$, dimension of mapped features: $d_z$.
\begin{algorithmic}[1]
\STATE \texttt{Third-party server do}\\
Generate random matrices $\mathbf{R}_A, \mathbf{R}_B, \mathbf{R}_b$, send $\mathbf{R}_A$ to A, and send $\mathbf{R}_B, \mathbf{R}_b$ to B;
\STATE \texttt{Client A do}
\STATE Calculate $\mathbf{X}_A^* =\overline{\mathbf{X}}_A+\mathbf{R}_A$, and send $\mathbf{X}_A^*$ to B;
\STATE \texttt{Client B do}
\STATE Generate random matrix $\mathbf{W}_B$, calculate $\mathbf{W}_B^* =\mathbf{W}_B+\mathbf{R}_B$, $\mathbf{E}_1 = \mathbf{X}_A^* \mathbf{W}_B+\mathbf{R}_b$, and send $\mathbf{W}_B^*, \mathbf{E}_1$ to A;
\STATE \texttt{Client A do}
\STATE Calculate $\mathbf{E}_2 = \mathbf{E}_1-\mathbf{R}_A \mathbf{W}_B^* = (\overline{\mathbf{X}}_A+\mathbf{R}_A)\mathbf{W}_B+\mathbf{R}_b-\mathbf{R}_A(\mathbf{W}_B+\mathbf{R}_B)=\overline{\mathbf{X}}_A \mathbf{W}_B+\mathbf{R}_b-\mathbf{R}_A \mathbf{R}_B$, and send $\mathbf{E}_2$ to Server;
\STATE \texttt{Third-party server do}
\STATE Calculate $\mathbf{E}_2-\mathbf{R}_b+\mathbf{R}_A \mathbf{R}_B=\overline{\mathbf{X}}_A \mathbf{W}_B$;
\STATE Repeat the above steps to calculate $\overline{\mathbf{X}}_B \mathbf{W}_A$ in the same interactive way;
\STATE \texttt{Client A do}
\STATE Calculate $\overline{\mathbf{X}}_A \mathbf{W}_A$ and send it to Server;
\STATE \texttt{Client B do}
\STATE Calculate $\overline{\mathbf{X}}_B \mathbf{W}_B$ and send it to Server;
\STATE \texttt{Third-party server do}
\STATE Generate random matrix $\mathbf{W}_1$, calculate $\mathbf{Z}^n$ using Eq.~\eqref{Z_n_4}.
%$
%\mathbf{Z}^n=\varphi
%\bigg(
%\left(\begin{array}{l}
%\overline{\mathbf{X}}_A \mathbf{W}_A ~~~\overline{\mathbf{X}}_A \mathbf{W}_B
%\overline{\mathbf{X}}_B \mathbf{W}_A ~~~\overline{\mathbf{X}}_B \mathbf{W}_B
%\end{array}\right) \times \mathbf{W}_1\bigg).
%$
\end{algorithmic}
{\bfseries Output:}  Mapped features $\mathbf{Z}^n$.%%\vskip -0.16in
\end{algorithm}

\subsection{The Complete Algorithm of MSBLS}\label{}
In this section, we give the pseudo code of MSBLS, as shown in Algorithm~\ref{MSBLS}. The complete MSBLS can be divided into three parts. The first part uses the data encryption and mapped feature generation procedures (Algorithm \ref{Interactive protocol of MSBLS}) to encrypt and fuse the data of the two clients. This process protects the privacy data of the two clients from being leaked and generates the mapped features required for the next step. The second part uses activation function $\xi(\cdot)$ and the mapped features to generate the enhancement features. In essence, this step is to expand the basis of the feature space to prepare for the calculation of weight coefficient in the next step. The third step uses the pseudo inverse to calculate the mapping relationship between the concatenation of the mapped features and the enhancement features and the output labels. In essence, this step is taking features as the basis to calculate the linear representation of the output label in the feature space.

\begin{algorithm}[!t]%%\vskip -0.06in
\caption{MSBLS.}
\label{MSBLS}
{\bfseries Input:} Data: Client A: $\mathbf{X}_A$, Client B: $\mathbf{X}_B$, feature mappings: $\varphi, \xi$, number of mapped features: $n$, dimension of mapped features: $d_z$, number of enhancement features: $m$, dimension of enhancement features: $d_h$, label of data $\mathbf{Y}\in\mathbb{R}^{N\times d_Y}$.
\begin{algorithmic}[l]
\STATE Calculate the mapped features $\mathbf{Z}^n$ via Algorithm \ref{Interactive protocol of MSBLS};
\STATE \texttt{Third-party server do}
\FOR {$j=1$ to $m$}
\STATE Generate random matrices $\mathbf{W}_{hj}$, $\bm{\beta}_{hj}$;
\STATE Calculate $\mathbf{H}_j = \xi (\mathbf{Z}^n \mathbf{W}_{hj} + \bm{\beta}_{hj})$;
\ENDFOR
\STATE Calculate $\mathbf{W}^m_n$ using Eqs.~\eqref{Y}-\eqref{A_pinv}.
\end{algorithmic}
{\bfseries Output:} Weight matrix $\mathbf{W}^m_n$.%%\vskip -0.16in
\end{algorithm}

\section{Security Analysis}\label{Security analysis}

In this section, we analyze the security of the protocol in Algorithm~\ref{Interactive protocol of MSBLS}, i.e. Protocol~\ref{Interactive protocol of MSBLS}. Specifically, we consider the data security in the semi-honest model. That is, the three parties (one server and two clients) involved in the calculation strictly implement Protocol~\ref{Interactive protocol of MSBLS}, but they will infer the non-held private data as much as possible according to the information they hold. In addition, we propose the following two assumptions.
\begin{itemize}
    \item Hypothesis 1. There is a secure channel between two of the three parties, that is, the data sent by the three parties will not be intercepted.

    \item Hypothesis 2. Two of the three parties will not collude with each other and share private data.
\end{itemize}

Security objective: on the premise of disclosing the mapped features, client A cannot obtain or recover data $\overline{\mathbf{X}}_B$ and $\mathbf{W}_B$, client B cannot obtain or recover data $\overline{\mathbf{X}}_A$ and $\mathbf{W}_A$, and the third-party server cannot obtain or recover $\overline{\mathbf{X}}_A, \overline{\mathbf{X}}_B, \mathbf{W}_A$ and $\mathbf{W}_B$.

\textbf{Security analysis:}

In line 4, client B receives $\mathbf{X}_A^*$, but because client B does not hold the matrix $\mathbf{R}_A$, it cannot recover $\overline{\mathbf{X}}_A$.

In line 6, client A receives $\mathbf{W}_B^*$ and $\mathbf{E}_1$, but since client A does not hold matrices $\mathbf{R}_B$ and $\mathbf{R}_b$, it cannot recover $\mathbf{W}_B$ and $\mathbf{X}_A^* \mathbf{W}_B$.

In line 8, client A obtains $\mathbf{E}_2$ through equation $\mathbf{E}_2 = \mathbf{E}_1-\mathbf{R}_A \mathbf{W}_B^*$, which is equivalent to $\overline{\mathbf{X}}_A \mathbf{W}_B+\mathbf{R}_b-\mathbf{R}_A \mathbf{R}_B$. Therefore, $\overline{\mathbf{X}}_A,\mathbf{R}_A$ and $\mathbf{E}_2$ are known quantities to client A but $\mathbf{R}_B$ and $\mathbf{R}_b$ are unknown quantities to client A, so it cannot recover $\mathbf{W}_B$.

In line 10, the third-party server obtains $\overline{\mathbf{X}}_A \mathbf{W}_B$ through equation $\mathbf{E}_2-\mathbf{R}_b+\mathbf{R}_A \mathbf{R}_B$. But $\mathbf{W}_B$ is unknown quantity to the third-party server, so it cannot recover $\overline{\mathbf{X}}_A$.

Line 11 is the same as the above analysis.

In lines 13 and 15, the third-party server holds the real values of the four product matrices $\overline{\mathbf{X}}_A \mathbf{W}_A$, $\overline{\mathbf{X}}_A \mathbf{W}_B$, $\overline{\mathbf{X}}_B \mathbf{W}_A$ and $\overline{\mathbf{X}}_B \mathbf{W}_B$, but using this information to recover the data $\overline{\mathbf{X}}_A$ and $\overline{\mathbf{X}}_B$ is an infeasible task. Specifically, the task is to solve a system of nonlinear equations of order $(N_A+N_B+n)\times d$ with $(N_A+N_B+n)\times d$ unknowns, which cannot be solved in polynomial time.

In line 17, the third-party server multiplies matrix $
\left(\begin{array}{l}
\overline{\mathbf{X}}_A \mathbf{W}_A ~~~\overline{\mathbf{X}}_A \mathbf{W}_B\\
\overline{\mathbf{X}}_B \mathbf{W}_A ~~~\overline{\mathbf{X}}_B \mathbf{W}_B
\end{array}\right)$ right by a random matrix $\mathbf{W}_1$. Since $\mathbf{W}_1$ is a non-public secret key, both clients A and B cannot recover matrix $
\left(\begin{array}{l}
\overline{\mathbf{X}}_A \mathbf{W}_A ~~~\overline{\mathbf{X}}_A \mathbf{W}_B\\
\overline{\mathbf{X}}_B \mathbf{W}_A ~~~\overline{\mathbf{X}}_B \mathbf{W}_B
\end{array}\right)$ using $\mathbf{Z}^n$.

It should be noted that in the parameter training process of the whole Algorithm~\ref{MSBLS}, Protocol~\ref{Interactive protocol of MSBLS} only runs once, and the corresponding data will only be transmitted once, which means that in theory, the data cannot be approximately recovered through the law of large numbers.

According to the above analysis, in the whole protocol procedure, neither client B nor the third-party server can recover the privacy data $\overline{\mathbf{X}}_A$ of client A. The same is true for $\overline{\mathbf{X}}_B$. Therefore, Protocol~\ref{Interactive protocol of MSBLS} is secure.

\section{Experiments}\label{sec: Experiments}
In this section, we verify the effectiveness of the proposed algorithm through experiments.

\subsection{Datesets Description and Experimental Scenario}\label{subsec: Datesets Description and Experimental Scenario}

In order to make the experimental results more convincing, this paper uses three classical image classification datasets, namely Norb, MNIST and Fashion.
\begin{itemize}
    \item The Norb dataset \cite{DBLP:conf/cvpr/LeCunHB04} samples the 3-D images of $50$ toys at different heights, angles and brightness. The images are divided into five categories: animals, humans, aircraft, trucks and cars. The training set contains $24300$ images of $25$ toys, while the testing set contains $24300$ images of another $25$ toys.

    \item The MNIST \cite{lecun1998gradient} handwritten numeral image dataset contains $70000$ scanned images of handwritten numerals $\{0,1,\dots,9\}$, each of which is a grayscale image of size $28 \times 28$. Among them, $60000$ samples are used as the training set and the other $10000$ samples are used as the testing set.

    \item The Fashion dataset \cite{DBLP:journals/corr/abs-1708-07747} is an upgraded version of the MNIST dataset. It contains $10$ categories of clothing images: T-shirt, trouser, pullover, dress, coat, sandal, shirt, snake, bag and ankle boot. The training set contains $60000$ samples with $6000$ samples per category, and the testing set contains $10000$ samples with $1000$ samples per category.
\end{itemize}

In order to simulate the actual scenario, we consider the distribution of the training data under the following two situations.
\begin{itemize}
    \item Quantity imbalance: the number of training samples held by each client is different. This situation is consistent with the actual scenario, because the amount of data held by different institutions (clients) is often different. In this experiment, we set the number of training samples of the two clients as six different ratios from $50\%:50\%$ to $5\%:95\%$.

    \item Non-IID: the label distribution of training data held by each client is not ``independent and identically distributed''. That is, the proportion of various labels of data held by each client is different. This situation is to simulate the difference of data samples of different institutions (clients) in the actual scene. Moreover, this experiment considers a more extreme and difficult scenario, that is, sorting the training data in the increasing order of class sizes, and then assigning the first half of the data to client A and the second half of the data to client B. This ensures that the labels of the data held by the two clients are almost different, which significantly increases the difficulty of model training.
\end{itemize}

\subsection{Baselines}

It should be noted that in MSBLS, the privacy of the testing data is also preserved. In addition, in order to verify the effectiveness of the algorithm, we design experiments from the following three perspectives:
\begin{itemize}
    \item By conducting comparison with the classical BLS \cite{DBLP:journals/tnn/ChenL18} without privacy protection, i.e., feeding the direct fusion of the training data from the two clients into the classical BLS, denoted as Non-privacy BLS, we will show that MSBLS does not lose the performance of the model on the premise of protecting data security.

    \item By conducting comparison with BLS in an absolute security environment, feeding the training data in each client into the classical BLS separately to train two independent BLS classifiers, denoted as Single-party BLS, we will show the performance difference between MSBLS and Single-party BLS, and hence draw a conclusion that MSBLS significantly outperforms Single-party BLS on the premise of protecting data privacy.

    \item By conducting comparison with the latest privacy protection machine learning method, namely federated learning algorithm (FedProx)~\cite{DBLP:conf/mlsys/LiSZSTS20}, we will show the performance difference between MSBLS and FedProx in terms of accuracy, training time and testing time.
\end{itemize}

\begin{table}[!t]
\caption{Comparison results of MSBLS and Non-privacy BLS (N-BLS) on the three datasets.}\label{MSBLS and BLS}
\resizebox{\columnwidth}{!}{
\begin{tabular}{@{}c|cc|cc|cc@{}}
\hline
Dataset & \multicolumn{2}{c|}{Norb} & \multicolumn{2}{c|}{MNIST} & \multicolumn{2}{c}{Fashion} \\ \hline
Methods      & MSBLS      & N-BLS         & MSBLS       & N-BLS         & MSBLS       & N-BLS          \\ \hline
Training accuracy    & 100\%       & 100\%       & 99.63\%      & 99.71\%     & 95.41\%      & 96.32\%      \\
Testing accuracy    & 88.38\%     & 88.12\%     & 98.51\%      & 98.54\%     & 89.81\%      & 90.16\%\\
\hline
\end{tabular}
}
\end{table}

\begin{table}[!t]
\caption{Comparison results of MSBLS and Single-party BLS (S-BLS) on the three datasets.}\label{MSBLS and single-party BLS}
\resizebox{\columnwidth}{!}{
\begin{tabular}{@{}c|cc|cc|cc@{}}
\hline
\multirow{2}{*}{\begin{tabular}[c]{@{}c@{}}Proportion of \\ training samples\end{tabular}} & \multicolumn{2}{c|}{Norb} & \multicolumn{2}{c|}{MNIST} & \multicolumn{2}{c}{Fashion} \\
\cline{2-7}
                                                                                           & MSBLS      & S-BLS    & MSBLS      & S-BLS     & MSBLS       & S-BLS     \\ \hline
50\%:50\%                                                                                        & 88.38\%     & 77.57\%     & 98.51\%     & 98.32\%      & 89.81\%      & 88.17\%      \\
40\%:60\%                                                                                        & 88.41\%     & 71.82\%     & 98.37\%     & 98.18\%      & 89.86\%      & 87.80\%      \\
30\%:70\%                                                                                        & 88.31\%     & 81.47\%     & 98.59\%     & 97.83\%      & 89.88\%      & 86.63\%      \\
20\%:80\%                                                                                        & 88.57\%     & 83.80\%     & 98.50\%     & 95.60\%      & 89.78\%      & 78.80\%      \\
10\%:90\%                                                                                       & 88.49\%     & 84.19\%     & 98.67\%     & 96.95\%      & 89.91\%      & 83.65\%      \\
5\%:95\%                                                                                   & 88.50\%     & 83.00\%     & 98.42\%     & 96.62\%      & 89.94\%      & 84.07\%\\
\hline
\end{tabular}
}
\end{table}

\subsection{Non-privacy Experiment Results}

Table \ref{MSBLS and BLS} reports the comparison results of MSBLS and Non-privacy BLS on the three datasets. The classification accuracy on both of the training dataset and the testing dataset is respectively reported. The results show that the performance difference between MSBLS and Non-privacy BLS is within $1\%$ on the premise of protecting data privacy, indicating that MSBLS will not lose the performance of the model while protecting data privacy. In fact, the mathematical forms of the mapped features and the enhancement features generated by MSBLS are consistent with those by Non-privacy BLS, so the final model performance will not be significantly different from that of Non-privacy BLS.

\begin{table*}[!t]
\caption{Comparison results of MSBLS and FedProx on the Norb dataset.}\label{MSBLS and FedProx(Norb)}
\centering
\begin{tabular}{c|cc|cc|cc}
\hline
\multirow{2}{*}{\begin{tabular}[c]{@{}c@{}}Proportion of\\ training samples\end{tabular}} & \multicolumn{2}{c|}{Training accuracy} & \multicolumn{2}{c|}{Testing accuracy} & \multicolumn{2}{c}{Training time (s)} \\
\cline{2-7}
                                                                                          & MSBLS            & FedProx           & MSBLS           & FedProx          & MSBLS          & FedProx         \\ \hline
50\%:50\%                                                                           & 100\%   & 100\%        & 88.38\%  & 88.14\%         & 29.35     & 386                   \\
40\%:60\%                                                                           & 100\%   & 100\%        & 88.41\%  & 87.65\%         & 29.06     & 389                   \\
30\%:70\%                                                                           & 100\%   & 100\%        & 88.31\%  & 87.00\%         & 29.49     & 384                   \\
20\%:80\%                                                                           & 100\%   & 100\%        & 88.57\%  & 87.28\%         & 28.95     & 404                   \\
10\%:90\%                                                                           & 100\%   & 100\%        & 88.49\%  & 87.29\%         & 27.66     & 427                   \\
5\%:95\%                                                                            & 100\%   & 100\%        & 88.50\%  & 87.37\%         & 28.82     & 410                   \\
\hline
\end{tabular}
\end{table*}

\begin{table*}[!t]
\caption{Comparison results of MSBLS and FedProx on the MNIST dataset.}\label{MSBLS and FedProx(Mnist)}
\centering
\begin{tabular}{c|ll|ll|lc}
\hline
\multirow{2}{*}{\begin{tabular}[c]{@{}c@{}}Proportion of\\ training samples\end{tabular}} & \multicolumn{2}{c|}{Training accuracy}                     & \multicolumn{2}{c|}{Testing accuracy}                       & \multicolumn{2}{c}{Training time (s)}    \\
              \cline{2-7}                                                                            & \multicolumn{1}{c}{MSBLS} & \multicolumn{1}{c|}{FedProx} & \multicolumn{1}{c}{MSBLS} & \multicolumn{1}{c|}{FedProx} & \multicolumn{1}{c}{MSBLS} & FedProx \\ \hline
50\%:50\%                                                                           & 99.63\%   & 99.27\%        & 98.51\%  & 98.72\%         & 53.79     & 304                   \\
40\%:60\%                                                                           & 99.72\%   & 99.26\%        & 98.37\%  & 98.85\%         & 52.47     & 318                   \\
30\%:70\%                                                                           & 99.61\%   & 99.74\%        & 98.59\%  & 98.02\%         & 52.14     & 314                   \\
20\%:80\%                                                                           & 99.77\%   & 99.44\%        & 98.50\%  & 98.41\%         & 53.36     & 337                   \\
10\%:90\%                                                                           & 99.53\%   & 99.30\%        & 98.67\%  & 98.72\%         & 54.33     & 339                   \\
5\%:95\%                                                                            & 99.58\%   & 99.02\%        & 98.42\%  & 98.88\%         & 52.12     & 352                   \\
\hline
\end{tabular}
\end{table*}

\begin{table*}[!t]
\caption{Comparison results of MSBLS and FedProx on the Fashion dataset.}\label{MSBLS and FedProx(Fashion)}
\centering
\begin{tabular}{c|ll|ll|lc}
\hline
\multirow{2}{*}{\begin{tabular}[c]{@{}c@{}}Proportion of\\ training samples\end{tabular}} & \multicolumn{2}{c|}{Training accuracy}                     & \multicolumn{2}{c|}{Testing accuracy}                       & \multicolumn{2}{c}{Training time (s)}    \\
\cline{2-7}                                                         & \multicolumn{1}{c}{MSBLS} & \multicolumn{1}{c|}{FedProx} & \multicolumn{1}{c}{MSBLS} & \multicolumn{1}{c|}{FedProx} & \multicolumn{1}{c}{MSBLS} & FedProx \\ \hline
50\%:50\%                                                                           & 95.41\%   & 93.42\%        & 89.81\%  & 86.80\%         & 66.27     & 296                   \\
40\%:60\%                                                                           & 96.24\%   & 93.88\%        & 89.86\%  & 87.01\%         & 65.73     & 308                   \\
30\%:70\%                                                                           & 95.39\%   & 93.20\%        & 89.88\%  & 86.62\%         & 66.91     & 318                   \\
20\%:80\%                                                                           & 95.78\%   & 92.43\%        & 89.78\%  & 86.05\%         & 67.41     & 317                   \\
10\%:90\%                                                                           & 96.05\%   & 93.09\%        & 89.91\%  & 85.59\%         & 67.50     & 332                   \\
5\%:95\%                                                                            & 95.87\%   & 92.17\%        & 89.94\%  & 85.37\%         & 66.33     & 334                   \\
\hline
\end{tabular}
\end{table*}

\subsection{Comparison of Single-party and Multi-party Experimental Results}

Table \ref{MSBLS and single-party BLS} reports the comparison results of MSBLS and Single-party BLS on the three datasets. The classification accuracy on the testing dataset is reported. The results show that when the proportion of the training samples in two clients changes, MSBLS always maintains a very stable testing accuracy on each dataset. On the contrary, the performance of Single-party BLS fluctuates significantly. This shows that the machine learning model trained independently will significantly lose the performance of the model. That is, when the proportions of data samples held by the two clients are different, if the two servers train the model parameters separately (i.e. Single-party BLS), it will be difficult for the client with a small number of samples to train a classifier with high accuracy, which is also an important reason why small hospitals need the help of large hospitals (on data). The use of MSBLS will not lose the accuracy of the model, so it can achieve secure privacy computing.

\subsection{Comparison with Federal Learning}

Table \ref{MSBLS and FedProx(Norb)}-\ref{MSBLS and FedProx(Fashion)} report the comparison results of MSBLS and FedProx on three datasets. Overall, compared with the cutting-edge federated learning algorithm, MSBLS has certain advantages in training accuracy and testing accuracy, and can save a lot of training time. This is because the performance of MSBLS is consistent with that of BLS, and the experimental effect of BLS has obvious advantages over the other machine learning algorithms such as deep neural network (see document \cite{DBLP:journals/tnn/ChenL18} for detailed comparison).

\subsection{Non-IID Scenario}

Table \ref{result_non-iid} reports the comparison results of MSBLS, Single-party BLS and FedProx in the Non-IID scenario. In this scenario, the testing accuracy of Single-party BLS is only about $50\%$, because the Single-party BLS trained alone can only classify part of the label data. For example, if the training set of client A does not contain samples with label $4$, the samples with label $4$ in the testing set can not be recognized at all. Both MSBLS and FedProx perform well in the Non-IID scenario. It is worth mentioning that MSBLS still has no performance loss in this scenario, which is due to the consistency between the theoretical results of MSBLS and the original BLS method.

\begin{table}[!t]
\caption{Experimental results of various algorithms in the Non-IID scenario.}\label{result_non-iid}
\centering
\begin{tabular}{c|c|c}
\hline
Dataset                  & algorithm & Testing accuracy \\ \hline
\multirow{3}{*}{Norb}    & MSBLS    & 88.19\%          \\ \cline{2-3}
                         & Single-party BLS  & 46.48\%          \\ \cline{2-3}
                         & FedProx        & 82.11\%          \\ \hline
\multirow{3}{*}{MNIST}   & MSBLS    & 98.61\%          \\ \cline{2-3}
                         & Single-party BLS  & 54.07\%          \\ \cline{2-3}
                         & FedProx        & 91.26\%          \\ \hline
\multirow{3}{*}{Fashion} & MSBLS    & 89.70\%          \\ \cline{2-3}
                         & Single-party BLS  & 46.95\%          \\ \cline{2-3}
                         & FedProx        & 81.04\%          \\ \hline
\end{tabular}
\end{table}

\section{Conclusions}\label{sec: Conclusions}

This paper proposes a new PPML method, which is a pioneering research work different from other methods. The existing PPML methods generally cannot simultaneously take into account multiple requirements such as security, application scope, efficiency and model performance. Specifically, differential privacy fails to simultaneously consider security and efficiency. The application scope and efficiency of homomorphic encryption are largely limited. The security of federal learning is not guaranteed by theory. The application scope of the traditional secure multi-party computing is relatively limited. The MSBLS method proposed in this paper inherits the advantages of both secure multi-party computing and neural network. It simultaneously takes into account the above four requirements, and achieves very satisfactory results (both in theory and experiment).

As described above, this paper has obtained a pioneering research achievement, which can provide an in-depth interactive perspective for the field of information security and machine learning. Researchers in these two fields can extend their research results to the field of PPML through the protocol provided in this paper. In other words, this paper opens up a new research path in the field of PPML, which combines the research results of the two fields more effectively. However, because this paper only makes a preliminary exploration, there are still some problems to be further studied. From the perspective of information security, first, different interactive protocols need to be designed in the scenarios of the semi-honest model and the malicious model. Secondly, we need to expand the number of clients participating in privacy computing from two to multiple. In this case, we need to consider the balance between security and communication times when designing interactive protocols. Third, how to ensure the security of data when some clients are attacked or colluded with each other? From the perspective of machine learning, firstly, MSBLS needs to adopt different feature extraction methods when applied to computer image, natural language, voice and other data types. Different feature extraction methods need to design corresponding interactive protocols to protect data security. Secondly, MSBLS needs to consider various practical needs, such as the security of stream data, the possible impact of missing data, data privacy calculation methods with different feature dimensions, etc.

%\section{References}

\bibliographystyle{IEEEtran}
\bibliography{ref}

\vskip -0.3in
\begin{IEEEbiography}
	[{\includegraphics[width=1in,height=1.25in,clip,keepaspectratio]{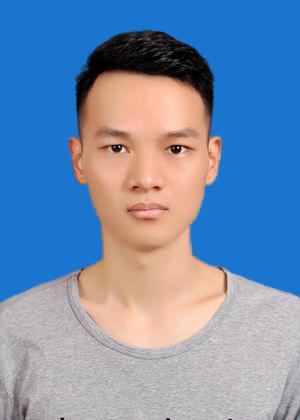}}]
	{Xiao-Kai Cao} received his Master degree in mathematics in 2020 from Guizhou University. He is purchasing his Ph.D. degree in computer science and technology at Sun Yat-sen University. His research interests are privacy computing and information security.
\end{IEEEbiography}
\vskip -0.3in
\begin{IEEEbiography}
	[{\includegraphics[width=1in,height=1.25in,clip,keepaspectratio]{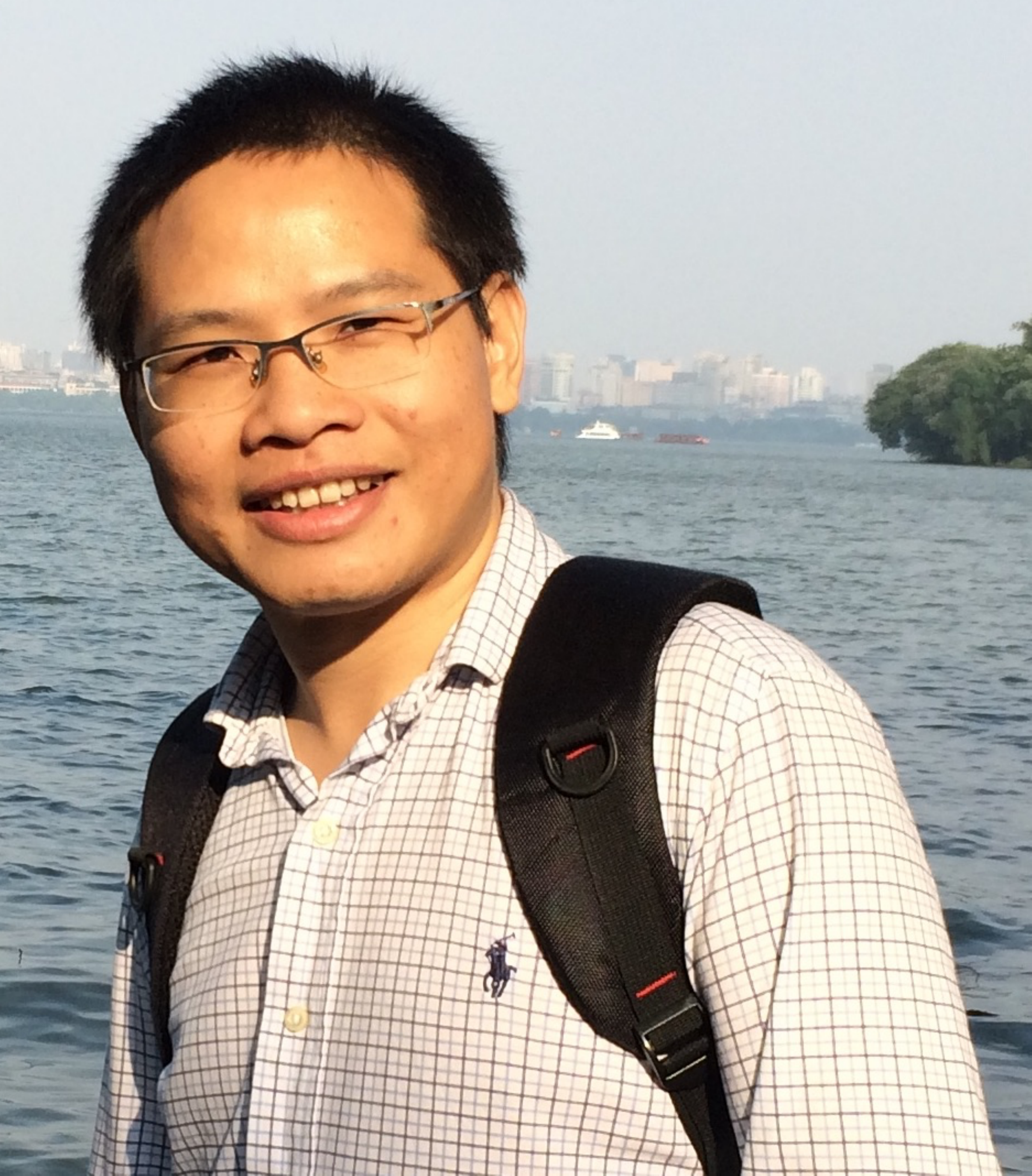}}]{Chang-Dong Wang} received the Ph.D. degree in computer science in 2013 from Sun Yat-sen University, Guangzhou, China. He was a visiting student at University of Illinois at Chicago from Jan. 2012 to Nov. 2012. He joined Sun Yat-sen University in 2013, where he is currently an associate professor with School of Computer Science and Engineering. His current research interests include machine learning and data mining. He has published over 80 scientific papers in international journals and conferences such as IEEE TPAMI, IEEE TKDE, IEEE TCYB, IEEE TNNLS, ACM TKDD, ACM TIST, IEEE TSMC-Systems, IEEE TII, IEEE TSMC-C, KDD, AAAI, IJCAI, CVPR, ICDM, CIKM and SDM. His ICDM 2010 paper won the Honorable Mention for Best Research Paper Awards. He won 2012 Microsoft Research Fellowship Nomination Award. He was awarded 2015 Chinese Association for Artificial Intelligence (CAAI) Outstanding Dissertation. He was an Associate Editor in Journal of Artificial Intelligence Research (JAIR).
\end{IEEEbiography}
\begin{IEEEbiography}
	[{\includegraphics[width=1in,height=1.25in,clip,keepaspectratio]{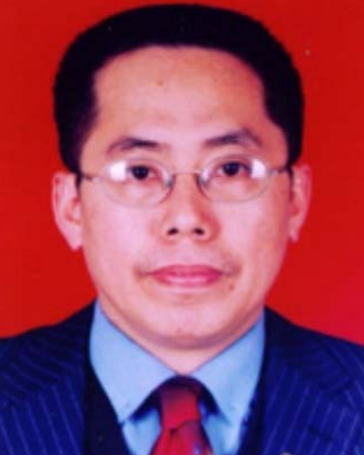}}]{Jian-Huang Lai} (Senior Member, IEEE) received the M.Sc. degree in applied mathematics and the Ph.D. degree in mathematics from Sun Yat-sen University, Guangzhou, China, in 1989 and 1999, respectively.

He joined Sun Yat-sen University, in 1989, as an Assistant Professor, where he is currently a Professor with the School of Data and Computer Science. He has authored or coauthored more than 200 scientific papers in the international journals and conferences on image processing and pattern recognition, such as IEEE TRANSACTIONS ON PATTERN ANALYSIS AND MACHINE INTELLIGENCE, IEEE TRANSACTIONS ON KNOWLEDGE AND DATA ENGINEERING, IEEE TRANSACTIONS ON NEURAL NETWORKS, IEEE TRANSACTIONS ON IMAGE PROCESSING, IEEE TRANSACTIONS ON SYSTEMS, MAN, AND CYBERNETICS-PART B: CYBERNETICS, Pattern Recognition, International Conference on Computer Vision, Conference on Computer Vision and Pattern Recognition, International Joint Conference on Artificial Intelligence, IEEE International Conference on Data Mining, and SIAM International Conference on Data Mining. His research interests include digital image processing, pattern recognition, and multimedia communication, wavelet and its applications. Dr. Lai is as a Standing Member of the Image and Graphics Association of China and also a Standing Director of the Image and Graphics Association of Guangdong.
\end{IEEEbiography}

\vskip -0.3in
\begin{IEEEbiography}
	[{\includegraphics[width=1in,height=1.25in,clip,keepaspectratio]{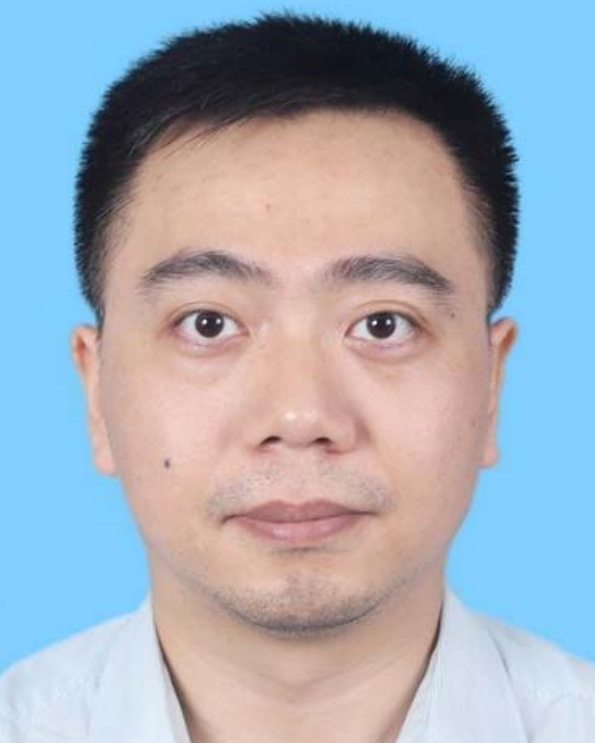}}]{Qiong Huang} (Member, IEEE) received the Ph.D. degree from the City University of Hong Kong in 2010. He is currently a Professor with the College of Mathematics and Informatics, South China Agricultural University, Guangzhou, China. He has authored or coauthored more than 120 research papers in international conferences and journals. His research interests include cryptography and information security, in particular, cryptographic protocols design and analysis. He was a Program Committee Member in many international conferences.
\end{IEEEbiography}
\vskip -2.8in
\begin{IEEEbiography}
	[{\includegraphics[width=1in,height=1.25in,clip,keepaspectratio]{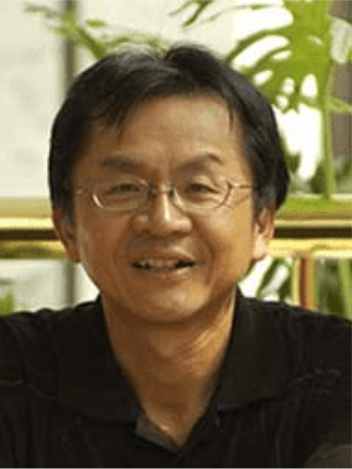}}]{C. L. Philip Chen} (Fellow, IEEE) received the M.S. degree from the University of Michigan at Ann Arbor, Ann Arbor, MI, USA, in 1985, and the Ph.D. degree from Purdue University, West Lafayette, IN, USA, in 1988, both in electrical and computer science.

He is currently the Chair Professor and the Dean of the College of Computer Science and Engineering, South China University of Technology, Guangzhou, China. Being a Program Evaluator of the Accreditation Board of Engineering and Technology Education in the U.S., for computer engineering, electrical engineering, and software engineering programs, he successfully architects the University of Macau's Engineering and Computer Science programs receiving accreditations from Washington/Seoul Accord through Hong Kong Institute of Engineers (HKIE), Hong Kong, of which is considered as his utmost contribution in engineering/computer science education for Macau as the Former Dean of the Faculty of Science and Technology. His current research interests include cybernetics, systems, and computational intelligence.

Dr. Chen was a recipient of the 2016 Outstanding Electrical and Computer Engineers Award from his alma mater, Purdue University, in 1988. He was also a recipient of the IEEE Norbert Wiener Award in 2018 for his contribution in systems and cybernetics, and machine learning. He is also a highly cited Researcher by Clarivate Analytics in 2018, 2019, and 2020. He is currently an Associate Editor for the IEEE TRANSACTIONS ON ARTIFICIAL INTELLIGENCE and IEEE TRANSACTIONS ON FUZZY SYSTEMS. He was the IEEE Systems, Man, and Cybernetics Society President from 2012 to 2013 and the Editor-in-Chief for the IEEE TRANSACTIONS ON CYBERNETICS and
 IEEE TRANSACTIONS ON SYSTEMS, MAN, AND CYBERNETICS: SYSTEMS. He was the Chair of TC 9.1 Economic and Business Systems of International Federation of Automatic Control from 2015 to 2017. He is a Fellow of AAAS, IAPR, CAA, and HKIE and a member of Academia Europaea, European Academy of Sciences and Arts.
\end{IEEEbiography}

\end{document}